\documentclass[amsmath,amssymb,pra,twocolumn]{revtex4}
\usepackage[cp1251]{inputenc}
\usepackage[english]{babel}
\usepackage{color}
\usepackage{graphicx}
\usepackage{dcolumn}
\usepackage{braket}
\usepackage{amsmath}
\usepackage{natbib}
\usepackage{hyperref}
\usepackage[T2A]{fontenc}
\usepackage{amsbsy}
\begin{document}
\title{Structured light pulses and their Lorentz-invariant mass}
\author{Stephen V. Vintskevich}
\affiliation{Moscow Institute of Physics and Technology,
Institutskii Per. 9, Dolgoprudny, Moscow Region 141700, Russia}
\author{Dmitry A. Grigoriev}
\affiliation{A.M. Prokhorov General Physics Institute,
Russian Academy of Sciences,38 Vavilov st., Moscow, 119991, Russia.}
\begin{abstract}
Lorentz-invariant mass and mean propagation velocity have been found for structured light pulses in vacuum considered as relativistic objects. We solved the boundary problem for such widely known field configurations as Gauss, Laguerre-Gauss, Bessel-Gauss, Hermite-Gauss, and Airy-Gauss ones. The pulses were taken having the finite duration temporal envelope. We discovered that Lorentz-invariant mass and mean propagation velocity significantly depend on the spatial-temporal structure of pulses. We found that mean propagation velocity is independent of the full energy of the pulse and is smaller than the speed of light.
\end{abstract}

\maketitle
\section{Introduction}
Infinitely extended in space and time electromagnetic plane waves propagate in a vacuum with the speed of light. But in reality electromagnetic fields have the form of light pulses with finite energy, duration, and occupy finite spatial volume. Moreover, light pulses can have a specific space-time structure, which can impart these pulses with unique properties.
Theoretical and experimental research of structured light pulses has been performed for decades. This field of optics is proved to be a fruitful area for practical applications in the fields of quantum information, spectroscopy, particles state control, etc. (for instance, see reviews \cite{Dunlop,STRUCT_LIGHT,Abramochkin} and reference therein).
Recently it was discussed that some specific types of structured light propagate in vacuum with mean propagation velocity smaller than the speed of light \cite{ALFANO,Sambles}. This conclusion was confirmed in various experiments involving quantum states of the light \cite{Giovannini}.
In work \cite{BOYD} authors emphasized that nontrivial configuration of spatial structure of the structured light leads to the deviation of its mean propagation velocity from the speed of light in vacuum $c$.
However, the problem of mean propagation velocity of light pulses in a vacuum can be attacked also from a different point of view. In accordance with Fourier analysis, a light pulse consists of a very large number of plane waves. One may associate these plane waves with photons. These plane waves propagate with the speed of light, $c$, but in different directions. Thus, the whole pulse is a result of interference of plane waves in superposition. Consequently, mean propagation velocity of pulses can be defined as a weighted superposition of velocities of individual plane waves. It was proved \cite{WE_2017_LP, WE_2017_EPL} that the defined in this way mean propagation velocity is smaller than the speed of light in vacuum. This approach is analogous to the problem of propagation of a group of particles in Special Theory of Relativity, including group of photons. In relativistic physics the mass of an arbitrary group of particles \cite{LL,OKUN1,OKUN2} is defined by the total energy of this group, $\varepsilon$, and its momentum, $\vec{p}$:
\begin{eqnarray}\label{MASS_INTRO}
m^2c^4 = \varepsilon^2 - c^2\vec{p}^2
\end{eqnarray}
Note, that for a single photon equation \eqref{MASS_INTRO} gives zero mass, as it has to be.
On the other hand, let us consider as an example a pair of photons with equal frequencies but non-collinear wave vectors. Their scalar product is given by $(\vec{k}_{1}\cdot\vec{k}_{2}) = \omega^2/c^2\cos(\theta), \theta \in \big(0, \pi \big]$. 
The total momentum of a photon pair is a vector with the absolute value smaller than total energy of the photon pair divided by $c$. In accordance with (\ref{MASS_INTRO}) the considered system of photons can be characterized by its Lorentz-invariant mass equal to
\begin{eqnarray}\label{MASS_SIMPLE}
m = \frac{2\hbar \omega}{c^2}\sin{\left(\frac{\theta}{2}\right)}
\end{eqnarray}\\
The physical meaning of Lorentz-invariant mass is related to the mean velocity of a group of particles as a whole, the absolute value of which is given by:
\begin{eqnarray}\label{VELOCITY}
v = \frac{c^2\lvert\vec{p}\rvert}{\varepsilon} = c\sqrt{1-\frac{m^2c^4}{\varepsilon^2}}
\end{eqnarray}
One can treat an arbitrary light pulse as a system consisting of photons, where a number of photons in the classical limit is very large.
Thus, one can apply the concepts of Lorentz-invariant mass and mean propagation velocity to arbitrary light pulses \cite{WE_2017_LP, WE_2017_EPL}.\\
In the present work, we calculate mean propagation velocity in accordance with Eq.(\ref{VELOCITY}) and Eq.(\ref{MASS_INTRO}) of localized light pulses specified by their spatio-temporal structure. In section (I.A), we start our analysis from most general perspective by defining relation between Lorentz-invariant mass and light pulse's electric field Fourier image. In the section (II.B), we define pulse's electric field component by setting the boundary problem. Boundary field is factorized on spatial and temporal functions which have the most general mathematical form. We completely describe pulses by their boundary problem using well-known methods of Fourier optics.
In the section (2) we specify structure of our pulses considering boundary conditions related with common types of structured light beams. It is worth to mention that all effects connected to mass and mean propagation velocity appear most clearly beyond the limits of paraxial approximation (see discussion in the section (III.B)). Thus, we perform numerical calculations of all pulse features for all considered examples. Based on these examples, we demonstrate that mean propagation velocity is smaller than $c$ for any considered pulse types and determined by its spatio-temporal structure.

\section{Lorentz-invariant mass of an arbitrary light pulse.}
\subsection{General description}
In this section, two problems are considered. First, we derive a general expression of Lorentz-invariant mass of light pulses in a vacuum.
Second, we discuss a propagation of an arbitrary pulse in the space-time domain, defined by its boundary problem for the electric field of the pulse. It will be demonstrated how the Fourier image of the boundary conditions can be used for calculation of Lorentz-invariant mass.\\
In works \cite{WE_2017_LP, WE_2017_CONF}, authors derived a method of calculation of Lorentz-invariant mass for the case of Gaussian light pulse propagation in a vacuum. It was shown that Lorentz-invariant mass is defined by a pulse spectrum
$\lvert\vec{E}\left(\vec{k}\right)\rvert$, where $\vec{E}\left(\vec{k}\right)$ is a Fourier image of a pulse electric field in the space-time domain:
\begin{eqnarray}\label{spectrum_k}
E\left(\vec{r},t\right) =\frac{1}{\left(2\pi\right)^\frac{3}{2}} \int\vec{E}\left(\vec{k}\right)e^{-i\vec{k}\vec{r}}e^{i\omega_{\vec{k}}t}d\vec{k},
\end{eqnarray}
where in accordance with Maxwell equations, $\omega_{\vec{k}} = c\sqrt{k_{z}^2+\vec{k}_{\perp}^2}$. In terms of Fourier images of $\vec{E}\left(\vec{k}\right)$ energy and momentum of a pulse are given by:
\begin{eqnarray}
&&\varepsilon = \frac{1}{2\pi}\int \lvert \vec{E}\left(\vec{k}\right) \rvert^2 d\vec{k}, \nonumber\\
&&\vec{p} = \frac{1}{2\pi c}\int \lvert \vec{E}\left(\vec{k}\right)\rvert^2 \frac{\vec{k}}{\lvert \vec{k} \rvert} d\vec{k}.
\end{eqnarray}
These equations and (\ref{MASS_INTRO}) yield the following general expression for  Lorentz - invariant mass:
\begin{eqnarray}\label{MASS_SPECTRUM_K}
&&m^2 = \frac{1}{\left(2\pi c^2\right )^2}\bigg[\left(\int \lvert \vec{E}\left(\vec{k}\right) \rvert^2 d\vec{k}\right)^2 \nonumber \\
&& - \left(\int \lvert \vec{E}\left(\vec{k}\right)\rvert^2 \frac{\vec{k}}{\lvert \vec{k} \rvert} d\vec{k}\right)^2\bigg]
\end{eqnarray}
At once, there is important family of beam models frequently used in optics of structured light. For this family of beams spatial spectrum possesses an axial symmetry with respect to the propagation axis.
Models having the axial symmetry are stemmed from Helmholtz equation in a paraxial limit in the cylindrical coordinates (e.g. see book \cite{SALEH}).
Spectra of such beams directly depend on the absolute value of transverse component of a wave vector
$\vec{E}\left(\vec{k}\right) =\vec{E}\left(k_{\perp},k_{z}\right)$.
It is evident that the mentioned symmetry of a spectrum $\vec{E}\left(\vec{k}\right)$  determines which component of the wave vector has the most significant contribution to the momentum calculation and, consequently, to the mass.
Thus, owing to the axial symmetry of the spectra, calculation of the mass is simplified, due to vanishing transverse components of the momentum in (\ref{MASS_SPECTRUM_K}):
\begin{eqnarray}\label{MASS_SPECTRUM_SIMPLE}
&&m^2 = \frac{1}{\left(2\pi c^2\right)^2}\bigg[\int\lvert \vec{E}\left(\vec{k}\right) \rvert^2 \left(1 - \frac{k_{z}}{\lvert k\rvert}\right)d\vec{k} \nonumber\\
&&\times \int\lvert \vec{E}\left(\vec{k}\right) \rvert^2 \left(1 + \frac{k_{z}}{\lvert k\rvert}\right)d\vec{k} \bigg]
\end{eqnarray}
\subsection{Boundary problem}
In previous section we derived most general form of Lorentz-invariant mass which is completely defined by pulse Fourier image $\vec{E}\left(\vec{k}\right)$ of a electric field. In turn, $\vec{E}\left(\vec{k}\right)$ can be specified by boundary conditions for electrical field transverse components. To complete pulse description one needs to derive $E_{z}$ component and magnetic field from Maxwell equations. In experiment, one can construct optical system with lenses, phase plates, nonlinear elements, etc. to give one or the other spatio-temporal structure. For instance, action of linear optical elements are especially straightforward in the Fourier space and reduced to algebraic transformation of field's Fourier image at some plane. We apply Fourier optics methods which are commonly used and proved as one of the most effective tools to describe the light pulse in space by its boundary problem (\cite{SALEH}).\\
Now let us turn to the boundary problem and specify its form.
Let pulse's electric field component at the plane $z = 0$ be defined as follows:
\begin{eqnarray}\label{Boundary problem}
\vec{E}_{\perp}\left(\vec{r}_{\perp},t,z=0\right)  = \vec{A}\left(x,y\right)T\left(t,\omega_0\right)
\end{eqnarray}
here $\vec{A}\left(x,y\right)$ is spatial part, $T\left(t, \omega_{0}\right)$ represents a temporal function depending on the carrier frequency $\omega_{0}$. We choose $T\left(t, \omega_{0}\right) = T\left(t\right)\sin\left(\omega_{0}t\right)$, where $T\left(t\right)$ is a real finite function (a temporal envelope) which commonly slowly varies with respect to highly oscillating $\sin\left(\omega_{0}t\right)$ function of carrier frequency.
Let us perform Fourier transform of (\ref{Boundary problem}) with Fourier images are considered in a ($\vec{k}_{\perp},\omega$) domain:
\begin{eqnarray}\label{FOURIER_TRANSFORM}
&&\vec{E}_{\perp}\left(\vec{k}_{\perp}, \omega\right)|_{z=0} = \int_{-\infty}^{\infty}\int_{-\infty}^{\infty}\vec{A}\left(x,y\right)e^{i\vec{k}_{\perp}\vec{r}}d\vec{r}_{\perp} \nonumber\\
&&\times\int_{-\infty}^{\infty} T(t) \sin\left(\omega_{0}t\right)e^{-i\omega t}dt
\end{eqnarray}
One commonly operates with the complex representation of electromagnetic filed. To obtain correct form of a field one needs to take a real part of a complex field vector. In order to presume correct form of a field in complex representation, the Fourier image (\ref{FOURIER_TRANSFORM}) should satisfy specific restrictions (\cite{SALEH}). The detailed discussion of imposed restrictions on Fourier image of temporal part is given in section (III.A). 
Now, our goal is to determine the field in the whole half-space $z\geq0$. We multiply the Fourier image of the boundary conditions $\vec{E}_{\perp}\left(\vec{k}_{\perp}, \omega\right)$ on the transfer function $h\left(\vec{k}_{\perp},\omega,z\right)$. Consequently, this Fourier expansion yields:
\begin{eqnarray}
h\left(\vec{k}_{\perp},\omega,z\right) = e^{i\sqrt{\frac{\omega^2}{c^2} - \vec{k}^2_{\perp}}z}
\end{eqnarray}
It worth to emphasize that it is especially useful to operate with Fourier images of field in $\vec{k}_{\perp}, \omega$ domain describing localized pulses, particularly in context of Lorentz-invariant mass calculation. Thus, we imply that $k_{z} = \sqrt{\omega^2/c^2 - {\vec{k}_{\perp}}^2}$ in accordance with the dispersion equation and fix the direction of the pulse propagation, $k_{z}>0$. Frequently, structured light pulses and beams are considered within the limits of paraxial approximation. In this case, one assumes that $\lvert\vec{k}\rvert \ll k_{z}$. 
In the present work we are not bounded by limits of paraxial approximation for obtained field in $z>0$. We will show that effects related to Lorentz-invariant mass and mean propagation velocity are appeared more clearly when divergence of a pulse is significant. That is equal to the fulfillment of $w_0\sim\lambda_0$, so small $w_0$ should be understood in this context as $w_0<50\lambda$. In this case our calculations can be performed only numerically.
Thereby, the resulting Fourier image of the pulse is:
\begin{eqnarray}\label{PROPAGATOR}
\vec{E}_{\perp}\left(\vec{k}_{\perp}, \omega\right) = h\left(\vec{k}_{\perp},\omega,z\right) \vec{E}\left(\vec{k}_{\perp}, \omega\right)|_{z=0}
\end{eqnarray}
Now we can use Maxwell equation $div\left(\vec{E}_{\vec{r},t}\right) = 0$ to find $E_{z}\left({\vec{k_{\perp}},\omega}\right)$ component of the field:
\begin{eqnarray}\label{Ez}
E_{z}\left({\vec{k_{\perp}},\omega}\right) =-\frac{k_{x}E_{x}\left(\vec{k}_{\perp}, \omega\right)+k_{y}E_{y}\left(\vec{k}_{\perp}, \omega\right)}{\sqrt{\frac{\omega^2}{c^2} - \vec{k}_{\perp}^2}},
\end{eqnarray}
This consequently gives $\vec{E}\left(\vec{k}_{\perp}, \omega\right)$.

At the third step, we restore the electromagnetic field in a space-time domain $z\geq 0$ by performing inverse Fourier transform in equation (\ref{PROPAGATOR}). It yields:
\begin{eqnarray}\label{final_field}
\vec{E}(\vec{r},t) = \frac{1}{({2\pi})^{\frac{3}{2}}}\int \vec{E}\left(\vec{k}_{\perp}, \omega\right)e^{-i\vec{k}_{\perp}\vec{r}}e^{i\omega t}d\vec{k}_{\perp}d\omega
\end{eqnarray}
In its turn, the magnetic field must be defined as follows
\begin{eqnarray}\label{Magnetic_FIELD}
\vec{H}\left(\vec{r},t\right) = \int\big[\frac{c\vec{k}\times\vec{E}\left(\vec{k}_{\perp}, \omega\right)}{\omega}\big]e^{-i\vec{k}_{\perp}\vec{r}}e^{i\omega t}d\vec{k}_{\perp}d\omega,
\end{eqnarray}
where $k_{z} = \sqrt{\omega^2/c^2 - \vec{k}_{\perp}^2}$.
On the next step we calculate the field for the $z\geq0$ and any time moment $t$ using $\vec{E}\left(\vec{k}_{\perp}, \omega\right)$ Fourier image, we
can obtain a connection with Fourier image $\vec{E}\left(\vec{k}_{\perp}, k_{z}\right)$. Owing to $k_{z} = \sqrt{\omega^2/c^2-\vec{k}_{\perp}^2}$, we make the variable change in integral (\ref{spectrum_k}):
\begin{eqnarray}\label{varable_change}
&&\vec{E}\left(\vec{r},t\right) =\frac{1}{\left(2\pi\right)^\frac{3}{2}}\int\frac{\omega\vec{E}\left(\vec{k}_{\perp},k_{z}\left(\omega\right)\right)}{c^2\sqrt{\omega^2/c^2-\vec{k}_{\perp}^2}}\nonumber\\
&&\times e^{-i\sqrt{\omega^2/c^2-\vec{k}_{\perp}^2}z}e^{-i\vec{k}_{\perp}\vec{r}}e^{i\omega t}d\vec{k}_{\perp}d\omega
\end{eqnarray}
At last, one can find the connection between spectra by direct comparison of the fields in (\ref{varable_change}) and (\ref{final_field}). As a result, we have:
\begin{eqnarray}\label{spectrum_comparison}
\vec{E}\left(\vec{k}_{\perp},k_{z}\left(\omega\right)\right) = \vec{E}\left(\vec{k}_{\perp}, \omega\right)\lvert_{z=0}\frac{c^2\sqrt{\omega^2/c^2-\vec{k}_{\perp}^2}}{\omega}
\end{eqnarray}
It permits one to get Lorentz-invariant mass from the Fourier image of the boundary conditions for various types of light pulses. One needs to calculate the Fourier image of the pulse's boundary condition and substitute it into the integral (\ref{MASS_SPECTRUM_K}) after the variable change similar to (\ref{varable_change}). Calculation of the mass has an especially elegant form in the case of pulses with the axial symmetry of the spectrum. It can be noticed in (\ref{MASS_SPECTRUM_SIMPLE}) that, in fact, this calculation can be reduced to the calculation of the following integral:
\begin{eqnarray}\label{INTEGRAL_SIMPLE}
p_{z,{\rm{sym}}} = c\int\rvert\vec{E}\left(\vec{k}_{\perp}, \omega\right)\lvert_{z=0}^2\left(1-\frac{\vec{k}_{\perp}^2c^2}{\omega^{2}}
\right)d\vec{k}_{\perp}d\omega
\end{eqnarray}
Consequently we can write:
\begin{eqnarray}\label{MASS_SIMPLE}
m = \frac{1}{c^2}\sqrt{\varepsilon^2 - \frac{c^2p_{z,{\rm{sym}}}^2}{4\pi^2}}
\end{eqnarray}

In the next section, we examine Lorentz-invariant mass and mean propagation speed of light pulses with well-known spatial profiles coincident with the Laguerre-Gauss, Bessel-Gauss, Hermite-Gauss beams at the boundary plane $z = 0$.


\section{Lorentz-invariant mass and mean propagation speed for the Gauss-family pulses.}
\subsection{Temporal part of the boundary conditions}
In \cite{WE_2017_LP, WE_2017_EPL, WE_2017_CONF} a model of a pulse with a Gaussian form of spatial and temporal envelopes was considered. It is a good approximation of the real pulses, especially in paraxial approximation, but formally such pulse has infinite duration. In contrast, here we adhere to model of the pulse which has finite duration in the time domain.
In accordance with (\ref{FOURIER_TRANSFORM}) Fourier image of temporal function reads
\begin{eqnarray}\label{FOURIER_TEMPORAL_image}
\tilde{S}\left(\omega\right) = \int_{-\infty}^{\infty} T\left(t\right)\sin\left(\omega_{0}t\right)e^{-i\omega t}dt
\end{eqnarray}
Note, that $T\left(t\right)\sin\left(\omega_{0}t\right)$ is the real function. It is imposed that $\tilde{S}\left(\omega\right)^{*} = \tilde{S}\left(-\omega\right)$.
Let us specify the  model of the temporal envelope $T\left(t\right)$ with finite duration. For simplicity, we choose:
\begin{eqnarray}
\label{STEP_F}
T\left(t\right)  =
\left\{
	\begin{array}{ll}
		1 & \mbox{if } t \in\left[0, t_{p} \right]\\
		0, & \mbox{otherwise}, 
	\end{array}
\right.
\end{eqnarray}
where, $t_{p}$ is a pulse duration.
In accordance with (\ref{FOURIER_TRANSFORM}), straightforward calculation of Fourier image of a temporal part reads:
\begin{eqnarray}\label{IMAGE_TEMPORAL}
&&\tilde{S}\left(\omega\right) =\frac{\left(-i\right)t_{p}}{2}e^{-\frac{i\omega t_{p}}{2}}\big[e^{\frac{i\omega_{0}t_{p}}{2}}{\rm{sinc}}\left(\frac{\left(\omega_{0}-\omega\right) t_{p}}{2}\right)\nonumber\\
&&-e^{\frac{-i\omega_{0}t_{p}}{2}}{\rm{sinc}}\left(\frac{\left(\omega_{0}+\omega\right) t_{p}}{2}\right)\big]
\end{eqnarray}
One needs to change Fourier image of temporal function of a field (\ref{IMAGE_TEMPORAL}) to operate correctly with the complex representation (or analytical signal)
as follows:
\begin{eqnarray}\label{ANANLYTICAL_ SIGNAL}
S\left(\omega\right)\longrightarrow
\left\{
	\begin{array}{ll}
		2\tilde{S}\left(\omega\right) & \mbox{if } \omega \geq 0\\
		0, & \mbox{otherwise} 
	\end{array}
\right.
\end{eqnarray}
The width of $S\left(\omega\right)$ is determined by pulse duration . We assume that parameter $\omega_{0}t_{p}\gg1$ i.e. the duration of pulses is quite big. Thus, one may assume that $S\left(\omega\right)$ has nonzero values at the interval with the center at $\omega_{0}$ and the width $\sim 1/t_{p}$.\\
Given model of the pulse's temporal part significantly differs from Gaussian envelope modulated with highly-oscillating function ${\rm{exp}}{\left(-\frac{t^2}{2t_p^2}\right)}\sin{\left(\omega_{0}t\right)}$, which was utilized previously in \cite{WE_2017_LP,WE_2017_CONF}.
The main difference is that the Fourier image of the latter has only imaginary part which is non-zero. In contrast, if a temporal envelope has the finite time duration then its Fourier image consists of an imaginary and real part. This fact does not significantly affect Lorentz-invariant mass calculation of a single coherent light pulse. However, as we believe, it can be important in the analysis of superposition of pulses, including partially-coherent and incoherent cases, as well as analysis of chirped pulses.
\subsection{Spatial part of the boundary conditions}
Once the temporal part is determined, we consider the spatial profile of the pulses. As a first example, let us consider pulses with axial symmetry. We specify pulse type by its spatial profile of the boundary conditions which are being chosen as the complex amplitude of Laguerre-Gauss (LG) and Bessel-Gauss (BG) beams at the plane $z = 0$, respectively:
\begin{eqnarray}\label{SPATIAL_COMPLEX_AMPLITUDE}
&&\vec{A}_{LG}\left(\rho\right)\rvert_{z=0} = C_{LG}\vec{e}_{\phi}\left(\frac{\rho_{\perp}}{w_{0}}\right)^{l}e^{-\frac{\rho^2}{w_{0}^2}}L_{q}^{l}\left(\frac{2\rho^2}{w_{0}^2}\right)e^{-il\phi}\nonumber\\
&&\vec{A}_{BG}\left(\rho\right)\rvert_{z=0} = C_{BG}\vec{e}_{\phi}J_{1}\left(\beta \rho\right)e^{-\frac{\rho^2}{w_{0}^2}},
\end{eqnarray}
where $C_{LG}$ and $C_{BG}$ are normalization constants
$\rho = \sqrt{x^2+y^2}$, $w_{0}$ is a pulse waist at the plane $z =0$, which is assumed to be the same for both types of pulses. $L_{l}^{q}\left(. \right)$ are generalized Laguerre polynomials and $l,m$ are positive integer numbers. We choose azimuthal polarization $\vec{e}_{\phi}$ for both cases of boundary conditions of pulses. It is a common choice for beams satisfying paraxial Helmholtz equation in the cylindrical coordinates \cite{zhan2009cylindrical,BARNET_ALLEN,HALL_VECTOR_BEAM,greene1998properties}.\\
Another important type of boundary conditions is stemmed from Hermite-Gauss and Airy-Gauss beams \cite{Siviloglou}.
\begin{eqnarray}
A_{HG} = \vec{e}_{45^{\circ}}C_{HG}\mathcal{H}_{l}\left(\frac{\sqrt{2}x}{w_{0}}\right)e^{-\frac{x^2}{w_{0}^2}}\mathcal{H}_{q}\left(\frac{\sqrt{2}y}{w_{0}}\right)e^{-\frac{y^2}{w_{0}^2}},
\end{eqnarray}
where $l,q$ are, again, positive integers. $\vec{e}_{45^{\circ}} = 1/\sqrt{2}\left(\vec{e}_{x}+\vec{e}_{y}\right)$ denotes linear polarization.
The case of Airy-Gauss is given by
\begin{eqnarray}
A_{AG} = \vec{e}_{45^{\circ}}C_{AG}\mathcal{A}i\left(\frac{x}{w_{0}}\right)e^{-\frac{x^2}{w_{0}^2}}\mathcal{A}i\left(\frac{y}{w_{0}}\right)e^{-\frac{y^2}{w_{0}^2}},
\end{eqnarray}
,where $\mathcal{A}i\left(. \right)$ - is Airy function.\\
Now, let us calculate Fourier images of spatial part of boundary problem performing Fourier transform with respect to the transverse spatial coordinates. After algebraic transformation and integration in the case of Laguerre-Gauss type we have:
\begin{eqnarray}\label{F_LG}
&&F_{LG}\left(k_{\perp}\right) = 2\pi C_{LG} w_{0}^2e^{i l\phi}e^{i\pi q}\left(\frac{k_{\perp}w_{0}}{2}\right)^l\nonumber \\
&&\times e^{-\frac{k_{\perp}^2w_{0}^2}{2}}L_{q}^{l}\left(\frac{k_{\perp}^2w_{0}^2}{2}\right)\vec{e}_{\phi}
\end{eqnarray}
Fourier transform in case of Bessel-Gauss pulse yields:
 \begin{eqnarray}\label{F_LG1}
\vec{F}_{BG}\left(k_{\perp}\right) = 2\pi C_{BG} w_{0}^2e^{-\frac{(\beta^2+\vec{k}_{\perp}^2}{2})}I_{1}\left(k_{\perp}\beta w_{0}\right)\vec{e}_{\phi}
\end{eqnarray}
\begin{figure}[h]
\centering
\includegraphics{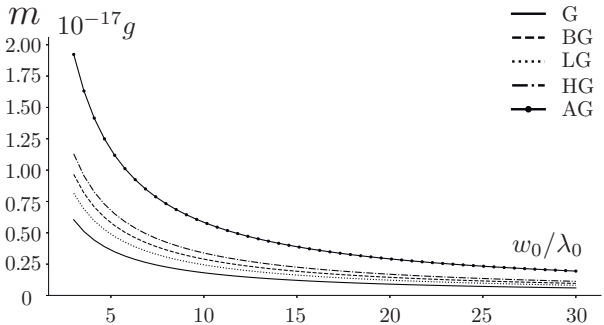}
\caption{Lorentz-invariant mass of the Gauss, Laguerre-Gauss ($l = 1, q = 0$), Bessel-Gauss, Hermite Gauss (symmetrical $l = 1,q = 1$) and Airy-Gauss light pulses. $w_{0}$ is a pulse waist at the plane $z = 0$, measured in numbers of the carrier wavelengths $\lambda_0 = 404\rm{\ nm}$ .
Numerical calculations were performed with respect to the fixed full energy $\varepsilon = 10 \ \rm{{mJ}}$ and pulse duration time $t_{p} = 0.539 \  \rm{{ps}} $ or $400$ periods of the carrier wave.}
\label{FIG1}
\end{figure}

\begin{figure}[h]
\centering
\includegraphics{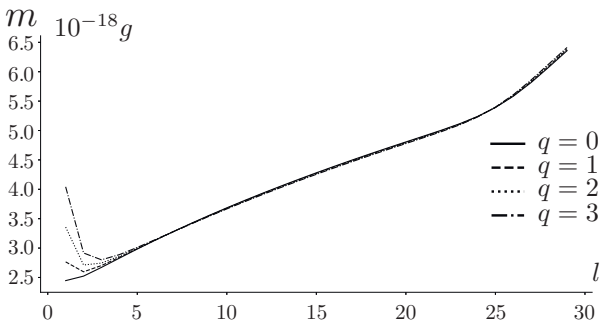}
\caption{Lorentz-invariant mass of the Laguerre-Gauss pulse as a function of the quantum number $l$, which corresponds to the orbital angular momentum. Different lines correspond to different parameters $q$.}
\label{FIG2}
\end{figure}

Finally, performing Fourier transform in case of the Hermite - Gauss pulse for arbitrary $l,q$ we have:
\begin{eqnarray}\label{F_HG}
&&F_{HG}\left(k_{x},k_{y}\right) =  \pi w_{0}^2i^{l+q}e^{\frac{-k_{\perp}^2w_{0}^2}{4}}\nonumber\\
&& \times \mathcal{H}_{l}\left(\frac{k_x w_{0}}{\sqrt{2}}\right)\mathcal{H}_{q}\left(\frac{k_y w_{0}}{\sqrt{2}}\right)
\end{eqnarray}
For Airy-Gauss type of pulse we perform only numerical culculations.
\begin{figure}
\centering
\includegraphics{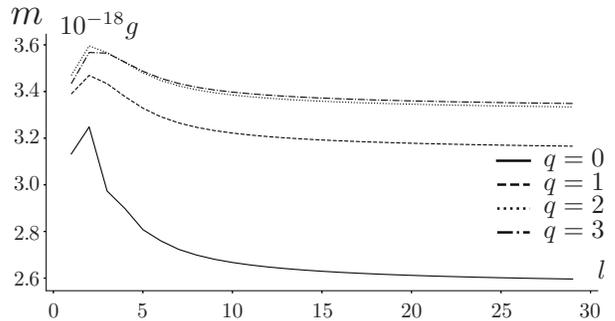}
\caption{Lorentz-invariant mass of Hermite - Gauss pulse as a function of number $l$ with different several parameters $q$. Energy, duration and wavelength are fixed and the same as in Figure (1,2). Pulse's waist is chosen to be $w_{0} = 10\lambda_0$.}
\label{FIG3}
\end{figure}


\begin{figure}
\centering
\includegraphics{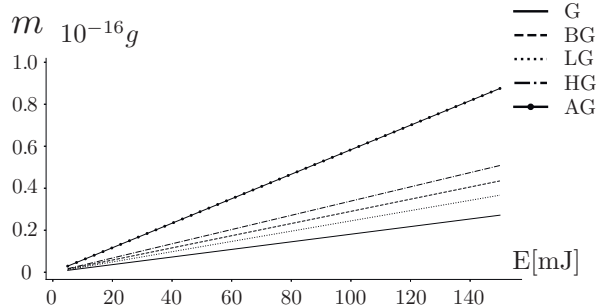}
\caption{Lorentz-invariant mass as a function of energy 
for different types. In this case duration, carrier wavelength, pulse's waist ($w_{0} = 10\lambda_0$) are fixed. The functions are linear for all types of pulses,  which indicates that velocity does not depend on energy and defined by structure of a pulse.
}
\label{FIG4}
\end{figure}



\begin{figure}[h]
\centering
\includegraphics{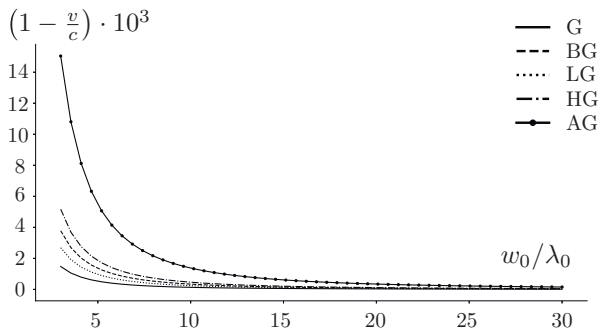}
\caption{Relative deviation of mean propagation velocity of Gauss, Laguerre-Gauss, Bessel-Gauss, Hermite-Gauss Airy-Gauss pulses, calculated by means of Lorentz-invariant mass \eqref{VELOCITY} as a function of a pulse waist. Other parameters are fixed.}
\label{FIG5}
\end{figure}


In our calculations of the spatial spectrum, we adopted several solutions from the book \cite{PRUDNIKOV} which contains integrals involving special functions of mathematical physics.\\
Different structures of pulses significantly influence on Lorentz-invariant mass and mean propagation speed of each type of a pulse. To demonstrate this fact, we have performed numerical calculations of integrals (\ref{MASS_SPECTRUM_K}) and (\ref{MASS_SPECTRUM_SIMPLE}) assuming that the total energy and time duration $t_{p}$ of the pulses are fixed. Therefore parameters of calculations are: wavelength $\lambda_0=404\rm{\ nm}$, waist $w_0=10\lambda_0$, time duration $t_{p}=400\lambda_0/c=0.539\rm{\ ps}$, central frequency $\omega_0=2\pi c/\lambda_0=4.7\cdot10^{15}\rm{\ Hz}$, and full energy $\varepsilon=10\rm{\ mJ}$. The results for the mass are presented in Fig.(\ref{FIG1}-\ref{FIG4}), and velocity dependencies are shown in Fig.(\ref{FIG5}). As can be seen, the structure of the pulse directly affects Lorentz-invariant mass, but these effects are non-negligible only in the case of the wide spatial spectrum.

\section{Conclusion}
In the present research Lorentz-invariant mass is calculated for structured light pulses in accordance with special relativity theory. Different pulses were described by different boundary conditions induced by beams, commonly used in practice and special temporal envelope, that provides the strict localization of pulses in the temporal domain. Based on general analytical derivations, all specific calculations were done numerically. The open-access Python source code for the calculations is available online \cite{GITHUB}. In all calculations and derivations the vector structure of the field was completely taken into account.
Obtained Lorentz-invariant mass appeared to be independent of the pulse duration at the given full energy of the pulse. On the other hand, Lorentz-invariant mass significantly depends on the waist of the pulse in all cases. Increasing of the waist leads to the decrease of Lorentz-invariant mass Fig(\ref{FIG1}). This is not surprising since the infinite waist corresponds to a plane wave, which mass equals zero. Thus, significant effects associated with non-zero Lorentz-invariant mass occur only at small $w_0$. For such $w_0$, the paraxial approximation is not valid, hence we perform the calculations numerically.
By means of Lorentz-invariant mass we found mean propagation velocity of the pulse. It behaves in some way similar to the mass. Mean propagation velocity is also independent of the pulse duration and phase of the temporal boundary conditions. It significantly depends on the waist of the pulse, but in contrast to the mass increases and tends to $c$ with the increase of the waist.
Linear dependencies in Figure(\ref{FIG4}) together with eq.\eqref{VELOCITY} indicates independence of mean propagation velocity of the pulse on its energy. Moreover, mean propagation velocity does not depend on the pulse duration, and as a result does not depend on the average power. Consequently, mean propagation velocity is fully defined by the space-time structure of the pulse.
In accordance with physical meaning, mean propagation velocity determines the rest frame of pulses. We believe that the structure of the field in its rest frame is an interesting subject. Finally in present work we consider coherent light pulses. The case of incoherent superposition of different light pulses within the context of Lorentz-invariant mass will be discussed elsewhere.

\section{Acknowledgement}
The work is supported by the Russian Foundation for Basic Research under Grant No. 18-32-00906. Authors thank Dr. Mikhail Fedorov and Viacheslav Sadykov for useful comments.


\begin{thebibliography}{20}
\providecommand{\natexlab}[1]{#1}
\providecommand{\url}[1]{\texttt{#1}}
\expandafter\ifx\csname urlstyle\endcsname\relax
  \providecommand{\doi}[1]{doi: #1}\else
  \providecommand{\doi}{doi: \begingroup \urlstyle{rm}\Url}\fi

\bibitem[Abramochkin and Volostnikov(2004)]{Abramochkin}
E.~G. Abramochkin and V.~G. Volostnikov.
\newblock Spiral light beams.
\newblock \emph{Usp. Fiz. Nauk}, 174\penalty0 (12):\penalty0 1273--1300, 2004.
\newblock \doi{10.3367/UFNr.0174.200412a.1273}.

\bibitem[Andrews"("2008")]{STRUCT_LIGHT}
"David~L. Andrews".
\newblock \emph{"Structured Light and Its Applications"}.
\newblock "Academic Press", "Burlington", "2008".
\newblock ISBN "978-0-12-374027-4".
\newblock \doi{"https://doi.org/10.1016/B978-0-12-374027-4.50003-6"}.

\bibitem[Rubinsztein-Dunlop et~al.(2016)Rubinsztein-Dunlop, Forbes, Berry,
  Dennis, Andrews, Mansuripur, Denz, Alpmann, Banzer, Bauer, Karimi, Marrucci,
  Padgett, Ritsch-Marte, Litchinitser, Bigelow, Rosales-Guzm{\'{a}}n, Belmonte,
  Torres, Neely, Baker, Gordon, Stilgoe, Romero, White, Fickler, Willner, Xie,
  McMorran, and Weiner]{Dunlop}
Halina Rubinsztein-Dunlop, Andrew Forbes, M~V Berry, M~R Dennis, David~L
  Andrews, Masud Mansuripur, Cornelia Denz, Christina Alpmann, Peter Banzer,
  Thomas Bauer, Ebrahim Karimi, Lorenzo Marrucci, Miles Padgett, Monika
  Ritsch-Marte, Natalia~M Litchinitser, Nicholas~P Bigelow,
  C~Rosales-Guzm{\'{a}}n, A~Belmonte, J~P Torres, Tyler~W Neely, Mark Baker,
  Reuven Gordon, Alexander~B Stilgoe, Jacquiline Romero, Andrew~G White, Robert
  Fickler, Alan~E Willner, Guodong Xie, Benjamin McMorran, and Andrew~M Weiner.
\newblock Roadmap on structured light.
\newblock \emph{Journal of Optics}, 19\penalty0 (1):\penalty0 013001, nov 2016.
\newblock \doi{10.1088/2040-8978/19/1/013001}.

\bibitem[Alfano and Nolan(2016)]{ALFANO}
Robert~R. Alfano and Daniel~A. Nolan.
\newblock Slowing of bessel light beam group velocity.
\newblock \emph{Optics Communications}, 361:\penalty0 25 -- 27, 2016.
\newblock ISSN 0030-4018.
\newblock \doi{https://doi.org/10.1016/j.optcom.2015.10.016}.

\bibitem[Sambles(2015)]{Sambles}
J.~R. Sambles.
\newblock Structured photons take it slow.
\newblock \emph{Science}, 347\penalty0 (6224):\penalty0 828--828, 2015.
\newblock ISSN 0036-8075.
\newblock \doi{10.1126/science.aaa6931}.

\bibitem[Giovannini et~al.(2015)Giovannini, Romero, Poto{\v c}ek, Ferenczi,
  Speirits, Barnett, Faccio, and Padgett]{Giovannini}
Daniel Giovannini, Jacquiline Romero, V{\'a}clav Poto{\v c}ek, Gergely
  Ferenczi, Fiona Speirits, Stephen~M. Barnett, Daniele Faccio, and Miles~J.
  Padgett.
\newblock Spatially structured photons that travel in free space slower than
  the speed of light.
\newblock \emph{Science}, 347\penalty0 (6224):\penalty0 857--860, 2015.
\newblock ISSN 0036-8075.
\newblock \doi{10.1126/science.aaa3035}.

\bibitem[Bouchard et~al.(2016)Bouchard, Harris, Mand, Boyd, and Karimi]{BOYD}
Fr\'{e}d\'{e}ric Bouchard, J\'{e}r\'{e}mie Harris, Harjaspreet Mand, Robert~W.
  Boyd, and Ebrahim Karimi.
\newblock Observation of subluminal twisted light in vacuum.
\newblock \emph{Optica}, 3\penalty0 (4):\penalty0 351--354, Apr 2016.
\newblock \doi{10.1364/OPTICA.3.000351}.

\bibitem[Fedorov and Vintskevich(2017{\natexlab{b}})]{WE_2017_LP}
M~V Fedorov and S~V Vintskevich.
\newblock Diverging light pulses in vacuum: Lorentz-invariant mass and mean
  propagation speed.
\newblock \emph{Laser Physics}, 27\penalty0 (3):\penalty0 036202, jan
  2017{\natexlab{b}}.
\newblock \doi{10.1088/1555-6611/aa567f}.

\bibitem[Fedorov et~al.(2017)Fedorov, Vintskevich, and Grigoriev]{WE_2017_EPL}
M.~V. Fedorov, S.~V. Vintskevich, and D.~A. Grigoriev.
\newblock Diffraction as a reason for slowing down light pulses in vacuum.
\newblock \emph{{EPL} (Europhysics Letters)}, 117\penalty0 (6):\penalty0 64001,
  mar 2017.
\newblock \doi{10.1209/0295-5075/117/64001}.

\bibitem[Landau(2013)]{LL}
Lev~Davidovich Landau.
\newblock \emph{The classical theory of fields}, volume~2.
\newblock Elsevier, 2013.

\bibitem[Okun\'(2000)]{OKUN1}
Lev~B Okun\'.
\newblock Reply to the letter 'What is mass?' by R.
  I. Khrapko.
\newblock \emph{Sov. Phys. Usp.}, 43\penalty0 (12):\penalty0 1270, 2000.

\bibitem[Okun(2009)]{OKUN2}
Lev~B Okun.
\newblock \emph{Energy and Mass in Relativity Theory}.
\newblock World Scientific, 2009.
\newblock \doi{10.1142/6833}.

\bibitem[Fedorov and Vintskevich(2017{\natexlab{a}})]{WE_2017_CONF}
M~V Fedorov and S~V Vintskevich.
\newblock Invariant mass and propagation speed of light pulses in vacuum.
\newblock \emph{Journal of Physics: Conference Series}, 826:\penalty0 012025,
  apr 2017{\natexlab{a}}.
\newblock \doi{10.1088/1742-6596/826/1/012025}.

\bibitem[Saleh and Teich(2007)]{SALEH}
Bahaa E~A Saleh and Malvin~Carl Teich.
\newblock \emph{{Fundamentals of photonics; 2nd ed.}}
\newblock Wiley series in pure and applied optics. Wiley, New York, NY, 2007.

\bibitem[Barnett and Allen(1994)]{BARNET_ALLEN}
Stephen~M Barnett and L~Allen.
\newblock Orbital angular momentum and nonparaxial light beams.
\newblock \emph{Optics communications}, 110\penalty0 (5-6):\penalty0 670--678,
  1994.

\bibitem[Greene and Hall(1998)]{greene1998properties}
Pamela~L Greene and Dennis~G Hall.
\newblock Properties and diffraction of vector bessel--gauss beams.
\newblock \emph{JOSA A}, 15\penalty0 (12):\penalty0 3020--3027, 1998.

\bibitem[Hall(1996)]{HALL_VECTOR_BEAM}
Dennis~G Hall.
\newblock Vector-beam solutions of maxwell’s wave equation.
\newblock \emph{Optics letters}, 21\penalty0 (1):\penalty0 9--11, 1996.

\bibitem[Zhan(2009)]{zhan2009cylindrical}
Qiwen Zhan.
\newblock Cylindrical vector beams: from mathematical concepts to applications.
\newblock \emph{Advances in Optics and Photonics}, 1\penalty0 (1):\penalty0
  1--57, 2009.

\bibitem{Siviloglou}
Georgios~A. Siviloglou and Demetrios~N. Christodoulides.
\newblock Accelerating finite energy airy beams.
\newblock {\em Opt. Lett.}, 32(8):979--981, Apr 2007.

\bibitem["Prudnikov et~al.("1983")"Prudnikov, Bry\v{c}kov, and
  Mari\v{c}ev]{PRUDNIKOV}
A.~P. "Prudnikov, Yu~A. Bry\v{c}kov, and O.~I." Mari\v{c}ev.
\newblock \emph{"Integrals and Series of Special Functions"}.
\newblock "Science", "Moscow, Russia", "1983".
\newblock ISBN "2-881-24682-6".

\bibitem[GitHub(2019)]{GITHUB}
Inc. GitHub.
\newblock Structured light pulses and their Lorentz-invariant mass.
  Source code for numerical calculations.
\newblock
  \href{https://github.com/GrigorievDmitry/Structured-light-pulses-and-their-Lorentz-invariant-mass.-Source-code-for-numerical-calculations}{https://github.com/GrigorievDmitry/Structured
  light pulses and their Lorentz-invariant mass. Source code for numerical
  calculations}, 2019.

\end{thebibliography}
\end{document}